# Amorphous intergranular films as toughening structural features

Zhiliang Pan, Timothy J. Rupert[*]

Department of Mechanical and Aerospace Engineering, University of California, Irvine, California 92697, USA

[*]To whom correspondence should be addressed: trupert@uci.edu

**Abstract**

The ability of amorphous intergranular films to mitigate damage formation at grain boundaries is studied with molecular dynamics simulations. We find that such films can alter both crack nucleation and crack growth rates by efficiently absorbing dislocations, with thicker films being more effective sinks. Local plastic strain brought by incoming dislocations is diffused into a triangular region within the amorphous film and is accommodated by a flow of boundary atoms which resembles a vortex shape; this vortex grows inside of the amorphous intergranular film as more dislocations are absorbed until it reaches the opposite amorphous-crystalline interface, after which cracks can finally be nucleated. Even after nucleation, these cracks grow more sluggishly in an amorphous intergranular film than they do along a clean grain boundary, since the driving force for crack growth is lower in the amorphous film. The results presented here suggest that amorphous intergranular films can act as toughening features within a microstructure, and thus are promising for designing nanostructured materials with better ductility and fracture toughness.

# 1. Introduction

Grain boundaries play an important role in the ductile failure of many polycrystalline materials by acting as preferential sites for damage nucleation [1-7]. During fatigue loading, interactions between dislocations lead to irreversible slip which accumulates during cycling, resulting in strain localization in the form of persistent slip bands [8] in both pure metals [9-11] and alloys [12]. Crack damage can be nucleated at the intersection between these slip bands and grain boundaries as a result of the dislocation pile-ups or local stress concentrations [13]. As another example, irradiation-assisted stress corrosion cracking [14-16] is caused by the formation of clear channels within the microstructure where subsequent dislocation movement is very easy [17, 18]. These channels usually terminated at grain boundaries, with this intersection being a prime site for potential damage formation [15]. Finally, grain boundary-dislocation interactions are also extremely important for the plastic deformation of nanocrystalline metals, where boundary sites act as nucleation sites [19, 20], pinning points [21], and absorption sites [22] for dislocation activity. In particular, absorption events should be very important for crack nucleation and failure. Bitzek et al. [23] used molecular dynamics (MD) simulations to show that the absorption of a single dislocation loop at a nanocrystalline grain boundary can lead to a pronounced increase of the local hydrostatic stress at the grain boundary, which should increase the driving force for intergranular fracture. Experimental work also supports this concept, with the in situ transmission electron microscopy experiments of Kumar et al. showing that grain boundaries are often the preferred sites for microcrack formation during loading [24].

Recent MD simulations [25] have shown that cracks can be formed at grain boundaries as a direct consequence of dislocation absorption when the plastic strain brought by the incoming dislocations towards the grain boundaries cannot be adequately accommodated through local rearrangement inside of the interface. As such, it is natural to propose that grain boundaries with strong ability to absorb incoming dislocations are desired to create materials with increased damage tolerance and fracture toughness. Van Swygenhoven and Derlet showed that preexisting free volume at grain boundaries can enhance the atomic shuffling events associated with grain boundary sliding [26]; similarly, preexisting excess free volume at a grain boundary can also make dislocation nucleation from the grain boundary easy to occur by decreasing the activation energy for such an event, as reported by Tucker and McDowell [27]. Based on these two observations, we hypothesize that additional preexisting free volume at grain boundaries might also enhance the atomic shuffling events associated with the dislocation absorption process.

If boundaries with a large amount of free volume are potentially able to delay damage nucleation, an amorphous interfacial phase may be a good candidate as a toughening structural feature. Amorphous materials are characterized by a lack of long-range crystalline order and contain excess free volume when compared to their crystalline counterparts [28]. Support for this idea that an amorphous phase can be beneficial has been provided by Wang and co-workers [29, 30], who made nanolaminates by separating nanocrystalline Cu layers with Cu-Zr amorphous intergranular films (AIFs). These materials exhibited superior tensile ductility when compared with nanocrystalline Cu alone, which the authors attributed to the ability of AIFs to act as high-capacity sinks for dislocations [29]. Wang and coauthors suggested that the fact that slip can be transferred to many directions in an amorphous material, unlike a crystalline phase where slip can only be transferred along specific close-packed directions, was responsible for this behavior. Brandl et al. [31] studied this model nanolaminate system with atomistic techniques, finding that the Peach-Koehler force posed by the elastically softer amorphous layer, as well as the interfacial shear of the amorphous-crystalline interface (ACI) itself, can attract dislocations and promote their absorption. It is notable that a number of other research groups have created a variety of nanolaminates by separating layers of amorphous material with crystalline/nanocrystalline Cu [32-38], crystalline Zr [39], or even amorphous phases with different properties [40]. However, the majority of these nanolaminates utilized the amorphous material as the matrix phase. In these studies, the amorphous metal is the majority phases and the primary goal is to suppress the catastrophic shear banding that leads to limited ductility in bulk metallic glasses [41, 42].

While empirical evidence suggests that AIFs can improve the ductility of nanolaminates, a complete physical description connecting grain boundary structure, dislocation absorption, and crack nucleation has not yet been provided. While previous studies have observed how a single dislocation can be absorbed [23], multiple absorption events and the accumulation of mechanical damage into full-fledged cracks has not been studied in detail. In this paper, we provide direct evidence of the toughening effect of AIFs, as well as physical insight into the mechanisms responsible for such behavior, using MD simulations. We find that AIFs significantly delay crack formation under an applied stress state that facilitates fracture, and that this toughening effect increases with increasing AIF thickness. The plastic strain brought by the incoming dislocations is shared through a wide, triangular-shaped region within the amorphous interface and accommodated through a vortex flow that grows inside the AIF, beginning from the ACI that intersects with incoming dislocations. Once plastic flow within the AIF reaches the opposite ACI, damage begins to accumulate and a crack can form. The rate of crack growth is also slowed by the introduction of an AIF. With the toughening effect of amorphous interfaces quantified, we also discuss strategies for designing better nanostructured materials using such interfaces.

## 2. Simulation Methods

MD simulations were performed with the Large-scale Atomic/Molecular Massively Parallel Simulator (LAMMPS) code [43] and all simulations used a 1 fs integration time step. Cu-Zr was chosen as a model alloy system here because robust interatomic potentials which recreate important properties of both crystalline Cu and amorphous Cu-Zr are available in the literature, and the aforementioned nanolaminate which inspired this work was also Cu-Zr. An embedded-atom method (EAM) potential was used to describe atomic interactions [44]. To produce a simulation cell that can systematically probe the effect of AIFs on dislocation accommodation and crack nucleation, we first created a reference bicrystal configuration of pure Cu with periodic boundary conditions applied in all three directions. This simulation cell is shown in Figure 1(a). The orientation of the center grain (G1) was chosen so that the resolved shear stress is highest on the horizontal slip plane and edge dislocations can be driven toward the boundaries. The second grain (G2) is oriented so that the resolved shear stresses on its slip planes are minimized. This means that the incoming dislocations are unlikely to be transmitted directly into G2, so that absorption at the grain boundary can be isolated.

Two slices were selected in the vicinity of the two grain boundaries and 25% of Cu atoms inside each slice were randomly replaced with Zr atoms. This composition (Cu-25 at.% Zr) is similar to the composition of the amorphous layer of nanocrystalline-amorphous nanolaminates [29] and is within the glass-forming range for Cu-Zr metallic glasses [31, 45]. To give these layers an amorphous structure, the atoms inside the selected slices were first melted at 1600 K and held for 200 ps while the rest of the atoms (i.e., the crystalline grain interiors) were held fixed. The melted atoms were then slowly quenched from 1600 K to 650 K over another 200 ps. To smoothly equilibrate the system, the constraint on the fixed atoms was removed and the entire system was quenched from 650 K to 10 K over 40 ps. The system was then kept at 10 K for an additional 20 ps to damp out any thermal fluctuations resulting from the termination of the quenching thermostat. To avoid unexpected thermal expansion, all of the above simulations were performed under the canonical ensemble where the volume of the simulation cell is fixed. Finally, to obtain a stress-free starting configuration, a Nose-Hoover thermo/barostat was used to relax the entire system at 10 K for 40 ps under zero pressure.

The radial distribution function of the center grain and an AIF region are shown in Figures 1(b) and (c), respectively. Figure 1(b) shows the sharp peaks expected from a crystalline solid, while Figure 1(c) shows the lack of long-range order in the interfacial region. The splitting of the second peak in Figure 1(c) is a sign that there is some short-range order in the glassy AIF. With this procedure, seven samples with grains of the same size and either clean grain boundaries or AIFs of different thickness (0.5 – 5.7 nm) were generated. It is important to note that we refer to a clean grain boundary as an AIF with zero thickness in some figures, to enable plotting of material response as a function of AIF thickness. Depending on the thickness of the AIF, the simulation cell is approximately 61- 73 nm long (X-direction), 32 nm tall (Y-direction), and 9 nm thick (Z-direction), containing ~1,400,000 to 1,700,000 atoms.

To investigate the damage resistance of an interface, it is necessary to create a positive hydrostatic stress to promote crack nucleation and growth in FCC metals [25, 46-49]. This was accomplished here by applying an elastic uniaxial tensile strain of 4% in the X-direction at a strain rate of $10^9$ $s^{-1}$, in a canonical ensemble. Poisson contraction was not allowed during this tension, resulting in a positive hydrostatic stress state in the sample. The chosen pre-strain (4%) in this work is adequate to promote crack formation at a grain boundary [25]. After this pre-tension step, each sample was equilibrated for 200 ps using the canonical ensemble to give extra time for boundary atoms to reach equilibrium. 10 starting configurations which are thermodynamically equivalent but differ slightly due to subtle thermal vibrations were created for each sample, in order to allow for increased statistics.

Finally, shear deformation under the canonical ensemble was applied to the seven interface types at engineering shear strain rates of $10^8$ s$^{-1}$ and $10^9$ s$^{-1}$. At the same time, an artificial dislocation source in the center of the sample was operated by gradually displacing two layers of atoms with respect to each other at a constant speed to generate dislocation pairs. The relative speed was chosen so that 10 pairs of parallel dislocations with opposite character on the same slip plane would be created by the time the global engineering shear strain reaches 20%. Our previous investigation of a clean grain boundary showed that any artificial or elevated stress state associated with this type of source is confined to a region far enough away from the interface and will not affect the interface-dislocation interactions [25]. During shear deformation, two layers of atoms at the bottom of the samples are held fixed in the vertical direction to limit rigid body grain rotation. Common neighbor analysis (CNA) was used to identify the local crystal structure of each atom [50], with FCC atoms colored green, hexagonal close packed (HCP) atoms red, body-centered cubic (BCC) atoms blue, and other atoms (usually grain boundary, dislocation, or crack surface atoms) white. All structural analysis and visualization of atomic configurations was performed using the open-source visualization tool OVITO [51].

## 3. Results

Figure 2 shows three samples with different grain boundary structures during the early stages of shear deformation at an applied strain rate of $10^9$ s$^{-1}$. Because Cu is an FCC metal with a low stacking fault energy, leading partial dislocations with stacking faults behind them are first emitted, followed by emission of the trailing partial dislocations. At an applied shear strain of 2%, the first complete dislocation pair has been produced by the artificial source and is beginning to propagate to the left and right. At a shear strain of 2.7%, the leading partial dislocations are absorbed by the two interfaces. At the same time, nucleation of the second dislocation pair is beginning in all samples. The three samples start to demonstrate different behavior at a shear strain of 4.3%. Although the second dislocation pair has been fully generated and propagates in all samples, Figure 2(a) shows that the trailing partials of the first dislocation pair have not been absorbed in the sample with clean grain boundaries. On the other hand, the first dislocation pair has been completely absorbed in the samples with AIFs, as shown in Figures 2(b) and (c), indicating that dislocations are more easily absorbed at an amorphous interface than at a clean grain boundary. This observation is consistent with the earlier predictions of Brandl et al. [31] based on observation of ACI structure.

The behavior of the three samples deviates even more clearly when additional shear strain is applied. Figure 3(a) shows that, when the applied shear strain reaches 6.0% and after the leading partial of the second dislocation pair is absorbed, the sample with clean grain boundaries nucleates a crack at the grain boundary-dislocation intersection on the right. The crack grows with the applied shear strain and finally the right grain boundary fractures completely at a shear strain of 8.5%. In contrast, in the sample with 1 nm thick AIFs shown in Figure 3(b), the first crack nucleates later at a shear strain of 7.8% and only a small crack through the sample thickness (along the Z-direction) is observed at the end of the shear deformation (applied shear strain of 20%). Both crack nucleation and growth are significantly delayed by the introduction of an amorphous interfacial phase. This toughening effect is further enhanced when increasing the AIF thickness to 3.8 nm. Figure 3(c) shows that crack nucleation occurs even later in this sample, at a shear strain of 9.5%. In addition, at the end of the shear deformation simulation we only find a roughly spherical crack embedded inside of the interface, instead of going all the way through the sample thickness as in the prior two cases.

While Figure 3 gives a qualitative sense of the toughening effect of AIFs, a more quantitative description can be found by identifying when cracks first form and then tracking their growth. To identify the damage which is a precursor to crack formation and then measure crack size, a very fine cubic mesh with a spacing of 0.2 Å was created near the two grain boundaries, following the work of Farkas et al. [52]. If a mesh point has no atoms sitting within 2.2 Å of it, this point is considered a potential defect and the sizes of individual defects are calculated by adding the volume associated with all connected mesh points also missing atoms. Although this fine mesh size and short search distance can detect defects as small as vacancies or even free volume in an amorphous region if necessary, we are interested in identifying damage sites which eventually evolve into a crack. A defect site is considered as "damage" if its volume is more than 0.014 $nm^3$, much larger than a free volume in amorphous region. As such, we do not count the free volume in the AIF as damage, since it is a necessary structural feature and instead focus on larger defects. The total volume of all damage in each grain boundary was tracked as a function of applied shear strain, as shown in Figure 4. Figure 4(a) shows that the amount of damage quickly increases in the clean grain boundary sample. However, when the clean grain boundary is replaced by a 1 nm thick AIF, the damage starts to accumulate much later and the damage volume grows much more slowly. Increasing the AIF thickness to 3.8 nm further suppresses the nucleation and growth of damage, as shown in Figure 4(c).

The critical strain for crack nucleation is identified by classifying a total damage volume greater than 0.5 $nm^3$, or the size of a spherical crack with diameter ~1 nm, as the first crack. Insets to Figure 4 show a zoomed view of the damage volume data, with this critical value for crack nucleation marked with a dotted line. Since the damage volume accumulates quickly at these early stages, we find that our measurements of critical strain are not overly sensitive to the exact damage volume used to define the nucleation event. The critical strain for crack nucleation for each simulation is plotted as a function of AIF thickness in Figure 5(a). The results show that the average critical strain, taken from ten identical starting configurations, increases with increasing AIF thickness at both strain rates. The average critical strain for the slower strain rate simulation set is lower than the higher strain rate data. Since interfacial fracture is related to dislocation absorption, we also plot the number of absorbed dislocations before crack nucleation as a function of the AIF thickness in Figure 5(b). This provides a more physical measurement of a boundary's ability to resist crack formation, and Figure 5(b) shows that the average number of the dislocations absorbed before crack nucleation also increases with the AIF thickness. In this case, strain rate has a negligible effect on the data.

The crack growth rate immediately after nucleation is also tracked to quantify the toughening effect. Growth rate was defined as the increase of crack volume per percent of shear strain in the interval where damage volume increases from 0.5 nm$^3$ to 5 nm$^3$. This interval was selected to focus on the early crack growth rate. Figure 6 shows the average crack growth rate plotted as a function of AIF thickness. The crack growth rate for the clean boundary is ~40 nm$^3$/% at strain rate of $10^9$ s$^{-1}$ and ~300 nm$^3$/% at strain rate of $10^8$ s$^{-1}$, far beyond the limit of the main vertical axis and shown in the inset to Figure 6. On the other hand, the average crack growth rates of the samples with AIFs are less than 10 nm$^3$/% for both strain rates. Inspection of Figure 6 shows that crack growth rate decreases with increasing AIF thickness, but tends to saturate after the AIF thickness reaches a critical value of ~4 nm. This suggests that there is a limit to the restriction that can be placed on crack growth, unlike the continual improvement of ability to resist crack nucleation observed in Figure 5.

## 4. Discussion

To understand how AIFs accommodate the strain heterogeneities brought by the incoming dislocations, Figure 7 shows the distribution of von Mises shear strain in samples with clean grain boundaries and 1 nm thick AIFs. At a shear strain of 6.0%, when crack nucleation starts in the clean grain boundary, the von Mises strain is highly concentrated inside the thin interface, as shown in Figure 7(a). The white dotted line in the close-up of the right dislocation-grain boundary intersection shows the relative shift between grains G1 and G2 across the grain boundary, indicating sliding of the grain boundary. However, at the same applied shear strain, when the clean grain boundary is replaced with 1 nm thick AIF in Figure 7(b), the von Mises strain is shared by the wider amorphous layer, especially in the region right ahead of the dislocation-ACI intersection, and the strain concentration is less severe. The white dotted line in the close-up shows that the relative shear displacement between grain G1 and G2 is also shared by the AIF. While the three layer offset from sliding is abrupt in the clean grain boundary sample, the AIF sample accommodates this offset gradually across a wider interface. This strain distribution/sharing process continues within a triangular region ahead of the dislocation-ACI intersection until an applied shear strain of 7.8%, when crack nucleation occurs. This observation suggests that the local plastic deformation brought by the incoming dislocations and strain concentration resulting from grain boundary sliding can be diffused into a wider region in the amorphous intergranular film. This process relieves the strain heterogeneity and, as a result, delays crack nucleation at the interface-dislocation intersection.

Figure 8 shows the local strain at a 5.7 nm thick AIF in a sample loaded at shear strain rate of $10^8$ s$^{-1}$. At a shear strain of 2%, the first leading partial dislocation is absorbed at the AIF, leading to plastic strain mainly at the ACI-dislocation intersection. However, a few additional atomic clusters in the AIF experience elevated strains as well, indicating the onset of shear transformation zones in these areas. At a shear strain of 4.5%, when the trailing partial of the first dislocation is also absorbed into the AIF, a roughly triangular region is formed right ahead of the ACI-dislocation intersection. Most of the strain is concentrated along upper boundary of the region. Increasing the shear strain to 6% does not increase the size or change the shape of the triangular region, but rather the atoms reach higher local strain levels inside this area. Finally, at shear strain of 10%, most of the triangular region is filled with atoms with von Mises strain larger than 0.3, as shown in Figure 8(d). In addition, interfacial sliding of the opposite ACI has been activated, as indicated by the strain concentration along the ACI on the left side of Figure 8(d).

To study the strain accommodation process from another perspective, the displacement field around an ACI-dislocation intersection of a sample loaded at shear strain rate of $10^8$ s$^{-1}$ is plotted in Figure 9. At an applied shear strain of 2%, after the first leading partial dislocation is absorbed, some atoms in the very vicinity of the ACI-dislocation intersection move a little to accommodate the first dislocation absorption event. In addition, some atoms away from the ACI-dislocation intersection in the amorphous region also move, again indicating the activation of shear transformation zones [53]. At a shear strain of 4.5%, the atoms beyond a triangular region, especially the upper section, move to accommodate the shift of the upper middle grain to the right. However, since the atoms largely shift together (i.e., there is no displacement gradient), this does not lead to high strain. The boundary between the shifted atoms and the stationary atoms roughly marks the boundary of the triangular region of high strain found in the previous figure. Large plastic strain is observed at the ACI due to the large displacement gradient there. Atoms close to the ACI-dislocation intersection starts to flow in a semi-circular pattern, resembling a vortex shape. The size of this vortex flow expands with increasing shear strain and number of absorbed dislocations (Figure 8(c)), and finally reaches the ACI on the left edge of the AIF at a shear strain of 10%, when a crack is nucleated at the ACI-dislocation intersection (Figure 8(d)). Note that, although a small amount of damage can be observed in Figure 8(c), it is much smaller than the critical value used to identify a crack nucleation event. When the shear strain is less than the critical strain characteristic of fracture at a clean grain boundary, the ACI itself can accommodate the plastic strain brought by the incoming dislocations; thereafter, the AIF takes over the accommodation process by forming a vortex flow around the ACI-dislocation intersection. More absorbed dislocations and larger shear strains lead to larger region of flow. When the boundary of the vortex flow reaches the opposite ACI, the growth of the flow is impeded and a crack can finally be nucleated.

In addition to delaying crack nucleation, AIFs were found to be advantageous for slowing crack growth. To isolate this result more clearly, Figure 10 presents a clean grain boundary sample where a crack forms and migrates. The crack is marked with red and viewed along the X-direction, so that the projection of the grain boundary plane is shown in the figure. The entire specimen cross-section is shown, but the dotted lines mark the limits of the mesh analysis "zone" around the grain boundary-dislocation intersection, marked with a dashed line. The crack nucleates at 5.91% shear strain and then grows in both the negative and positive Z-directions (left and right, respectively), as well as in the negative Y-direction (down). The crack has grown through the entire thickness of the sample by an applied shear strain of 6.2%. Thereafter, the crack grows in the negative Y-direction until it reaches the lower boundary of the mesh used for the analysis at an applied shear strain of 6.8%. At shear strain of 7.1%, after the crack reaches the lower boundary of the simulation cell, it begins to grow in the positive Y-direction and quickly reaches the upper boundary of the mesh at shear strain of 8%. The entire grain boundary fractures shortly after at shear strain of 8.4%. In contrast to the rapid growth at clean grain boundaries, the crack grows much slower at AIFs. As shown in Figure 11 for a 5.7 nm thick AIF right at the crack nucleation point, four damage sites can be observed along the dislocation-ACI intersection line. These damage sites grow slowly and remain restricted to a local region until, at shear strain of 16.8%, the smallest crack (colored grey in the first image) coalesces with another crack (colored green). The remaining three cracks keep growing slowly and no more coalescence events occur through the end of the deformation simulation.

To understand why cracks grow so rapidly in clean grain boundaries but slowly in AIFs, the atomic stress distributions from the two interfaces shown in the first frames of Figures 10 and 11 (i.e., at the crack nucleation event) are presented in Figure 12. The viewing angle is the same as Figures 10 and 11 and the dislocation-interface intersections are marked with dashed lines. Only atoms in the interfaces are visualized and the cracks are outlined with solid white lines. Figure 12(a) shows the hydrostatic stress distribution at the clean grain boundary, which is higher below the dislocation-grain boundary intersection than above, except for at the crack surface. Slightly elevated hydrostatic stress can also be observed directly around the crack. These observations are consistent with the findings shown in Figure 10, where the crack grows outwards as well as downward. This suggests that hydrostatic stress is the driving force for crack growth and the crack is propagated in an opening mode (i.e., Mode I). In contrast, the von Mises stress below the dislocation-grain boundary intersection is lower than that above the intersection, as shown in Figure 12(b). Figures 12(c) and (d) show the atomic hydrostatic stress and von Mises stress distributions, respectively, inside ACI shown in Figure 11. The stress is distributed more randomly inside the ACI instead of concentrating around the cracks, indicating that the driving force for crack growth inside ACI is much lower. This lower local hydrostatic stress gives rise to the much slower crack growth rate inside the AIF, as compared to the clean grain boundary.

The finding that grain boundaries with planned structural disorder can both delay crack nucleation and slow crack propagation has the potential to enable the design of nanostructured materials with improved mechanical properties. Nanocrystalline metals and alloys are infamous for having low ductility [54], which limits their usefulness in spite of their excellent strength. However, control of boundary structure opens a pathway for avoiding this limitation. For nanocrystalline grain sizes above ~10 nm, plasticity remains dominated by dislocation-based mechanisms [55] and dislocation pile-ups at grain boundaries can even still occur [56]. The vast majority of thermally-stable nanocrystalline alloys which have been produced (see, e.g., [57-60]) exist in this grain size regime, making strategies for dislocation accommodation extremely important. If clean grain boundaries are replaced by AIFs, then dislocations will be less likely to pile-up since they can be efficient absorbed. Farkas, et al. [61] showed that crack growth in nanocrystalline metals can be dominated by the coalescence of microvoids at the grain boundaries. Our results, specifically those shown in Figure 11, demonstrate that the small cracks formed inside of AIFs grow extremely slowly with increased levels of deformation, remain restricted to a local region, and do not coalesce even after high levels of applied shear strain. The nanocrystalline-amorphous Cu-Zr nanolaminates produced by Wang et al. [29] demonstrated the power of this concept by showing higher ductility when compared to nanocrystalline Cu films, as discussed in the introduction to this paper. However, these materials, while extremely instructive as a model system, have their drawbacks when searching for a general materials design strategy. First, nanolaminate systems are predestined to have highly anisotropic properties due to the layer-by-layer deposition technique used. While AIF planes are located periodically through the film thickness, there are no amorphous interfaces in other directions. In addition, deposition techniques such as sputtering are not amenable to the production of bulk quantities of material. A better design principal would enable the production of bulk nanostructured materials with good properties in all directions.

The concept of grain boundary "complexions," or thermodynamically-stable interfacial phases, should offer an opportunity for designing interface structure. While grain boundary premelting was suggested many years ago [62], complexions in single element systems are extremely rare [63]. However, recent research has found that complexions are much more common in multicomponent systems, such as $ZnO-Bi_2O_3$ [64], $Al_2O_3-CaO$, and $Al_2O_3-SiO_2$ [65]. Dillon et al. [65] recently helped formalize the complexion concept further, by identifying six distinct interfacial structures, ranging from lightly doped boundaries to wetting disordered films, in alumina and connecting these structures to different GB mobilities. While the majority of observed complexions have been in ceramic systems, Luo et al. [66] found AIFs in W-Ni and suggested that such films were responsible for the previously unexplained phenomenon of activated sintering. The complexion concept should make it possible to introduce AIFs into nanostructured materials through grain boundary segregation [67], but this has not be extended to nanostructured materials yet. The recent work of Murdoch and Schuh [68] makes it possible to identify stable nanocrystalline alloys where strong segregation of dopant elements occurs, potentially enabling such grain boundary design. Because AIFs would be distributed in random orientations if introduced into nanocrystalline materials, the anisotropy limitation of a nanolaminate would not occur. In addition, the techniques used to induce complexion transitions in polycrystalline materials have been simple, traditional techniques such as doping followed by heat treatment [69, 70]. These procedures do not preclude the fabrication of bulk materials. In fact, a recent review by Tschopp et al. [71] identifies a number of processing techniques that enable bulk nanostructured materials where complexion engineering could easily be integrated.

Another important design parameter identified by our study which has practical implications is the thickness of the AIF. Our data shows that thicker amorphous interfaces are better for toughening, although the effect on crack growth rates saturates. It is expected that AIF thickness should also play an important role in bulk materials with randomly distributed amorphous complexions. However, current theoretical descriptions of grain boundary phase diagrams [70], while extremely useful for identifying the temperature and compositions which promote formation of amorphous complexions, treat all grain boundaries as random high angle interfaces with one common boundary energy. In real random polycrystals, there will be a wide variety of grain boundary character, with an associated range of grain boundary energies. This means that there will likely be a distribution of AIF thicknesses in even the best designed alloy system. For example, low energy twin boundaries may not transform into a higher level complexion even at very high temperatures near the solidus temperature. In addition, AIFs which are too thick might even degrade toughness. If one can think of traditional polycrystalline materials as having zero thickness AIFs, then a material with AIFs thicker than the average grain size would be a traditional metallic glass. Even slightly below this last extreme would be a material with an amorphous metal as its majority phase (i.e., a bulk metallic glass-based composite with embedded crystalline particles). If there is a continuous amorphous phase through a macroscopic sample, catastrophic shear banding can occur directly through the glassy material. Therefore, there is likely a limit to how thick AIFs should be before they degrade ductility. Although these are still open questions, the modulation of AIF thickness through intelligent doping will provide a wider design space for nanocrystalline materials.

## 5. Conclusions

In this work, we performed MD simulations to show that AIFs can serve as a toughening structural features. Our results show that AIFs can delay crack nucleation by efficiently absorbing incoming dislocations. The strain concentration brought by incoming dislocations is diffused into a triangular region within the AIF and accommodated through a vortex flow of boundary atoms around the ACI-dislocation intersection. The size of this vortex flow grows with the increasing number of absorbed dislocations, until it reaches the opposite ACI. At this point, the flow is saturated and crack nucleation occurs. As such, the ability of an AIF to delay crack formation from dislocation absorption increases with increasing AIF thickness, since a larger area for cooperative flow exists. AIFs can also suppress crack growth after nucleation, by alleviating stress concentrations around interfacial cracks. This work provides direct evidence that AIFs can act as tough interfaces, and these interfacial structures should be promising structural features for the design of tough nanocrystalline materials.

## Acknowledgements

This research was supported by the U.S. Army Research Office through Grant W911NF-12-1-0511.

**Figures and captions**

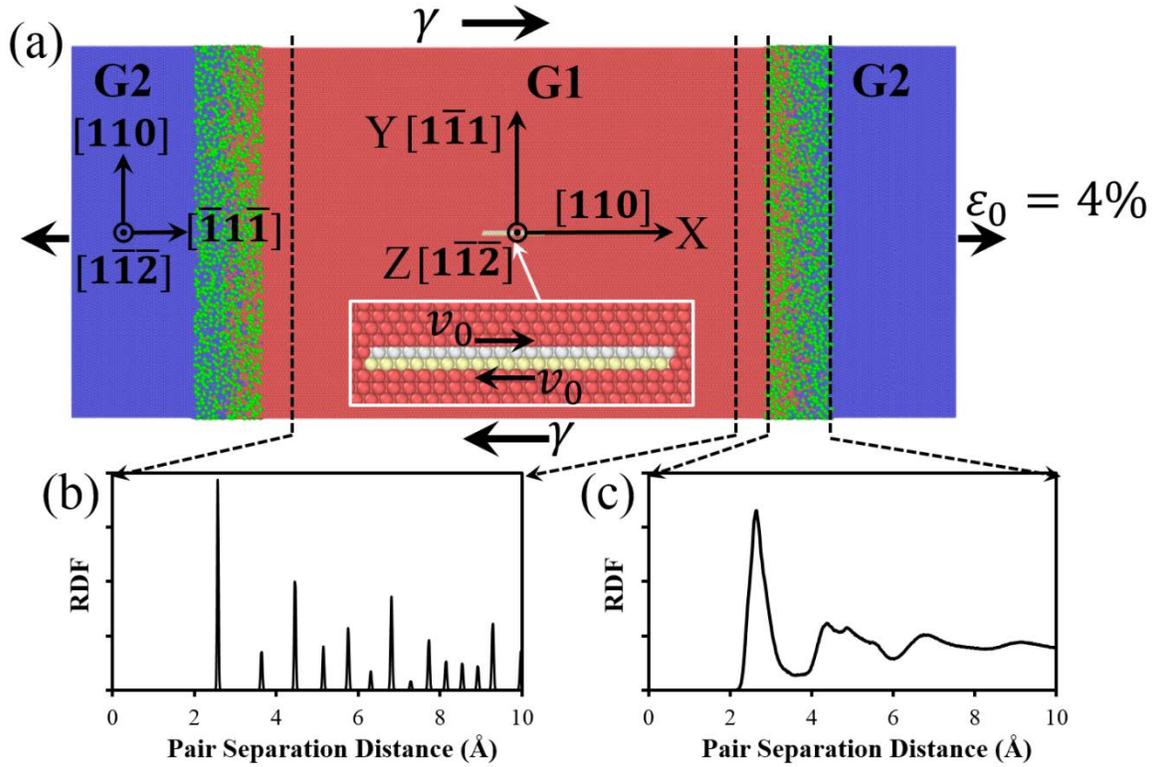

**Figure 1. (a) Bicrystal sample containing grains G1 (red atoms) and G2 (blue atoms), as well as AIFs doped with Zr (green atoms). Inset shows a dislocation source created in the center of G1 by moving the yellow and white atoms relative to each other at a constant speed. Radial distribution functions for grain G1 (b) and one AIF (c) are also shown here.**

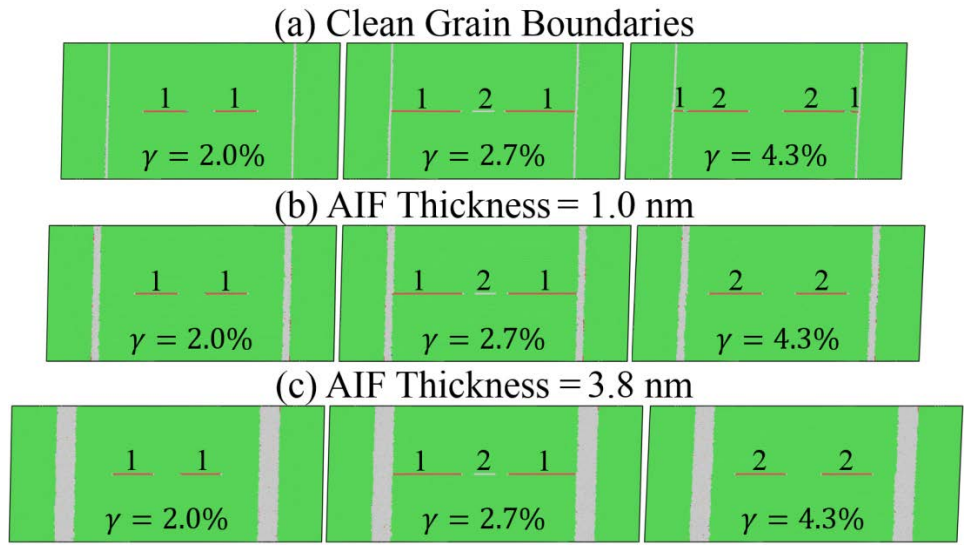

**Figure 2.** Dislocation emission and absorption observed during shear deformation at a strain rate of $10^9$ s$^{-1}$ in samples with (a) clean grain boundaries, (b) 1 nm thick AIFs, and (c) 3.8 nm thick AIFs. The natural numbers label the sequence of dislocations generated from the source in the center of the specimen. Snapshots at shear strain of 4.3% clearly demonstrate that dislocations are more easily absorbed into AIFs than into clean grain boundaries.

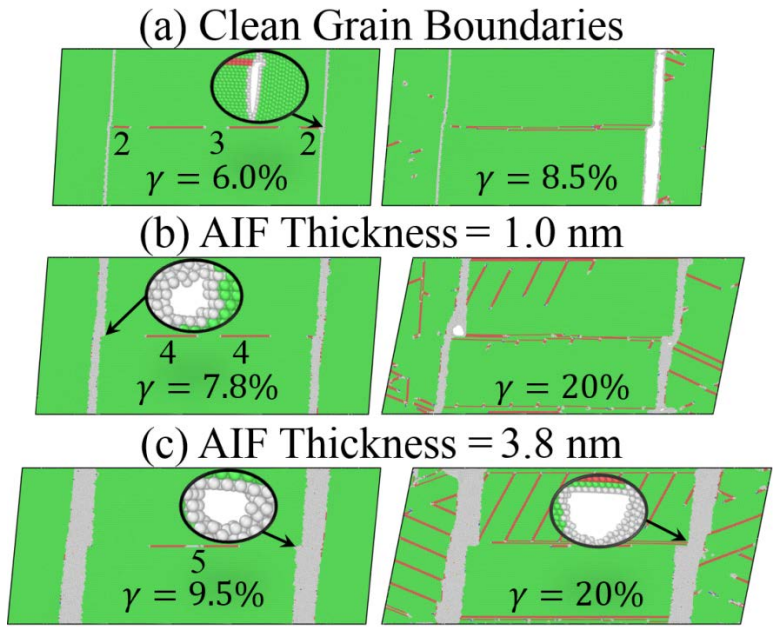

**Figure 3.** Crack nucleation and growth observed during shear deformation at strain rate of $10^9$ s$^{-1}$ in samples with (a) clean grain boundaries, (b) 1 nm thick AIFs, and (c) 3.8 nm thick AIFs. Both crack nucleation and growth are suppressed in AIFs. The natural numbers label the sequence of dislocation pairs.

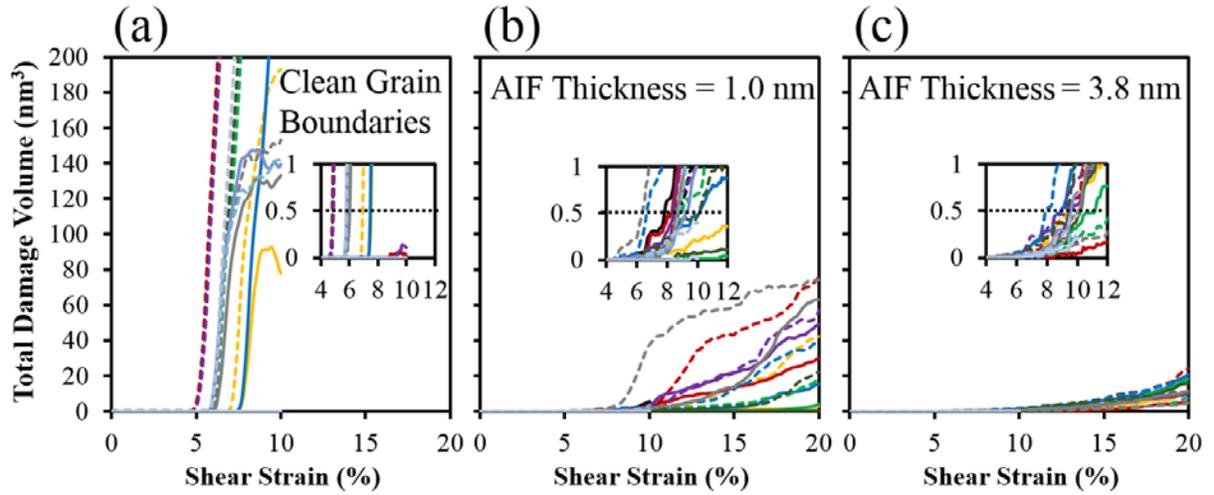

**Figure 4.** Damage volume as a function of shear strain at strain rate of $10^9$ s$^{-1}$ for samples with (a) clean grain boundaries, (b) 1 nm thick AIFs, and (c) 3.8 nm thick AIFs. Curves are colored according to starting configurations. Solid curves are for the left grain boundary, while the dashed for the right. Insets show close-ups of the crack evolution curves during the nucleation stage, where the dotted black lines mark the position of 0.5 nm$^3$ of damage, or the crack nucleation event.

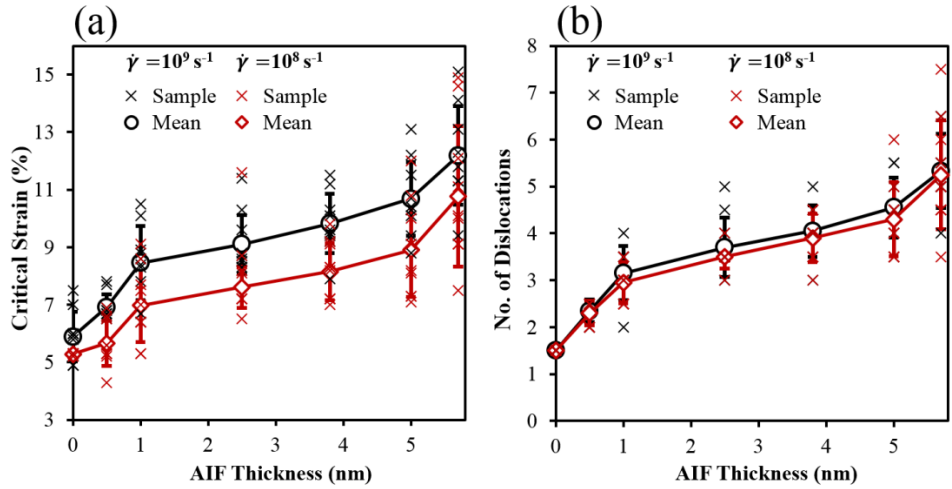

**Figure 5.** (a) Critical strain for crack nucleation and (b) number of dislocations absorbed before crack nucleation as a function of AIF thickness. The trend lines show that thickening of the AIF delays crack nucleation.

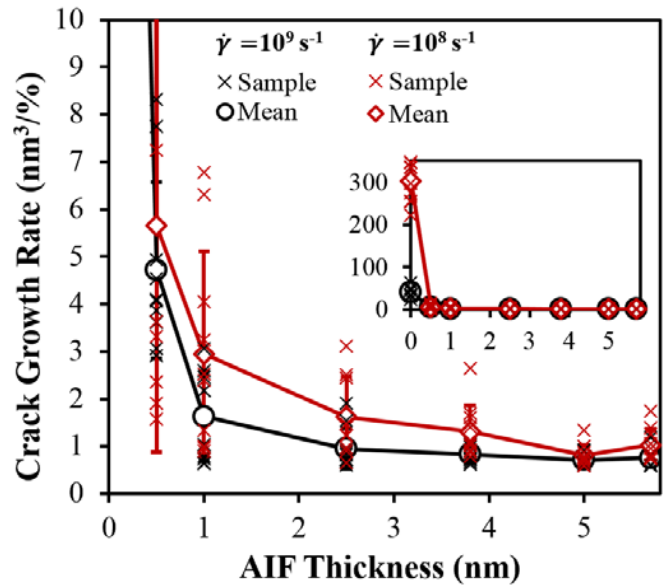

**Figure 6.** Crack growth rate, from 0.5 nm$^3$ to 5 nm$^3$ to isolate initial growth rates, as a function of AIF thickness. Inset is the same figure with larger maximum bounds in the vertical axis so that data points from clean grain boundaries can also be presented.

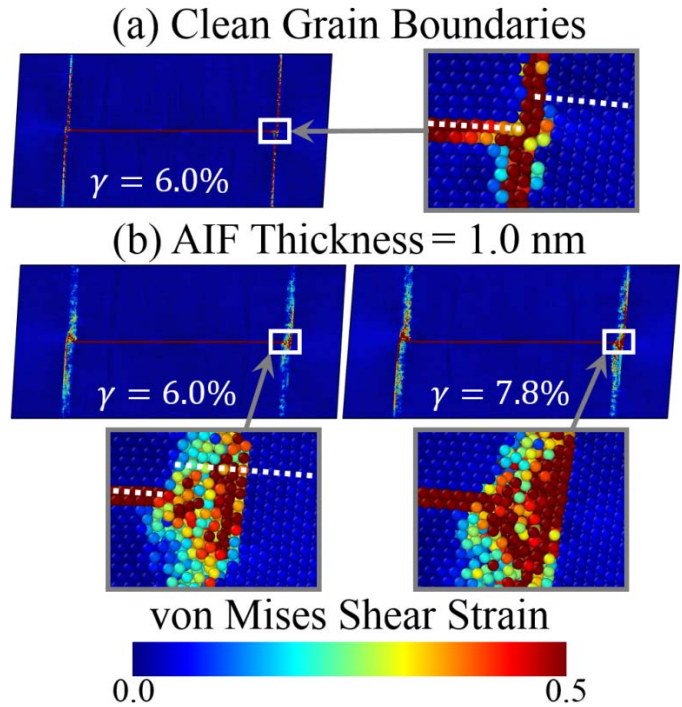

**Figure 7.** The distribution of atomic von Mises shear strain in samples with (a) clean grain boundaries and (b) 1 nm thick AIFs. White dotted lines in the close-ups show the relative shear displacement between grain G1 and G2 across the clean grain boundary or AIF.

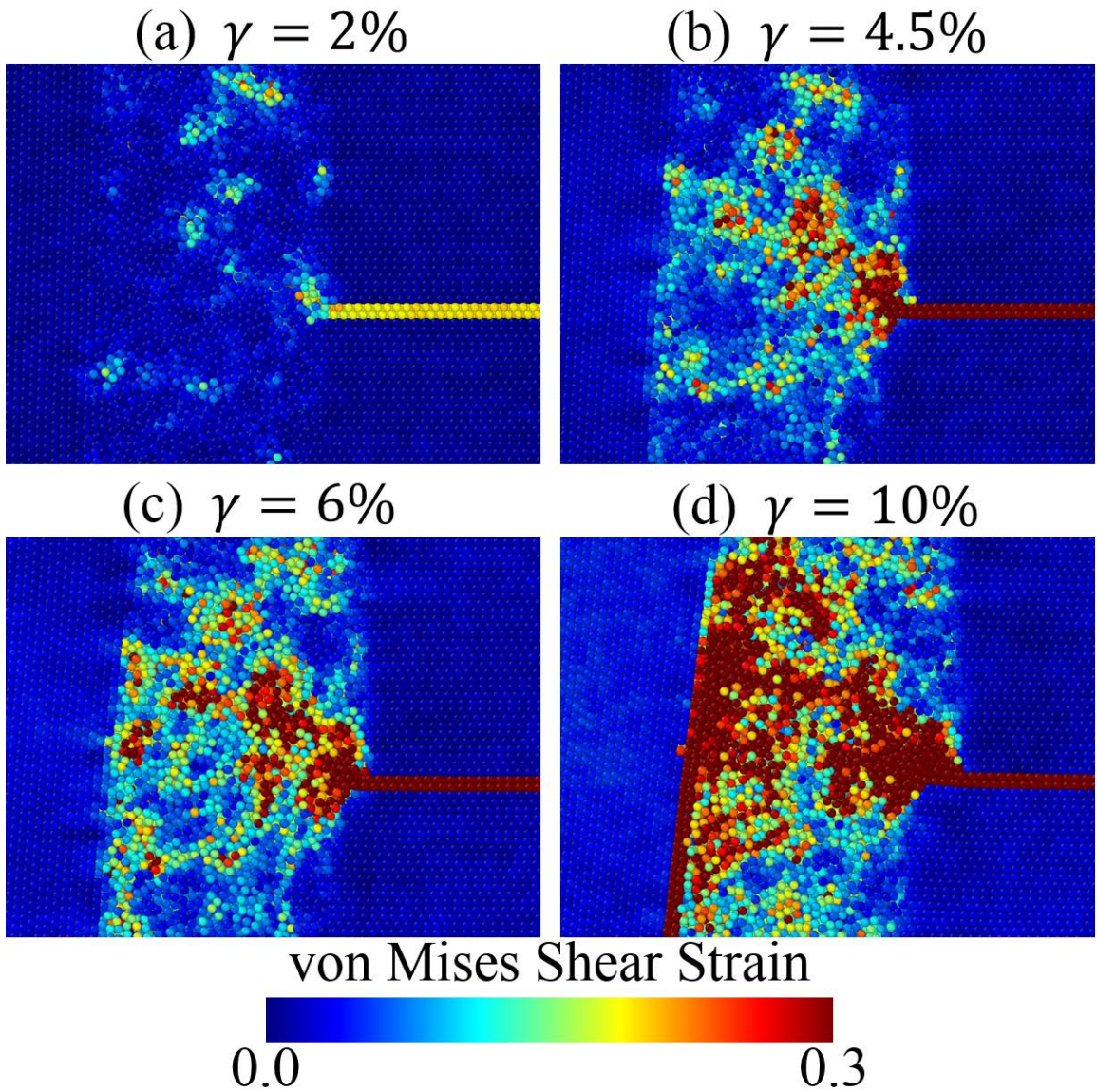

Figure 8. The distribution of atomic von Mises strain at a 5.7 nm thick AIF in a sample at shear strains of (a) 2%, (b) 4.5%, (c) 6%, and (d) 10%, when loaded at shear strain rate of $10^8$ s$^{-1}$. A triangular region of high strain gradually forms at the AIF with increasing applied shear strain.

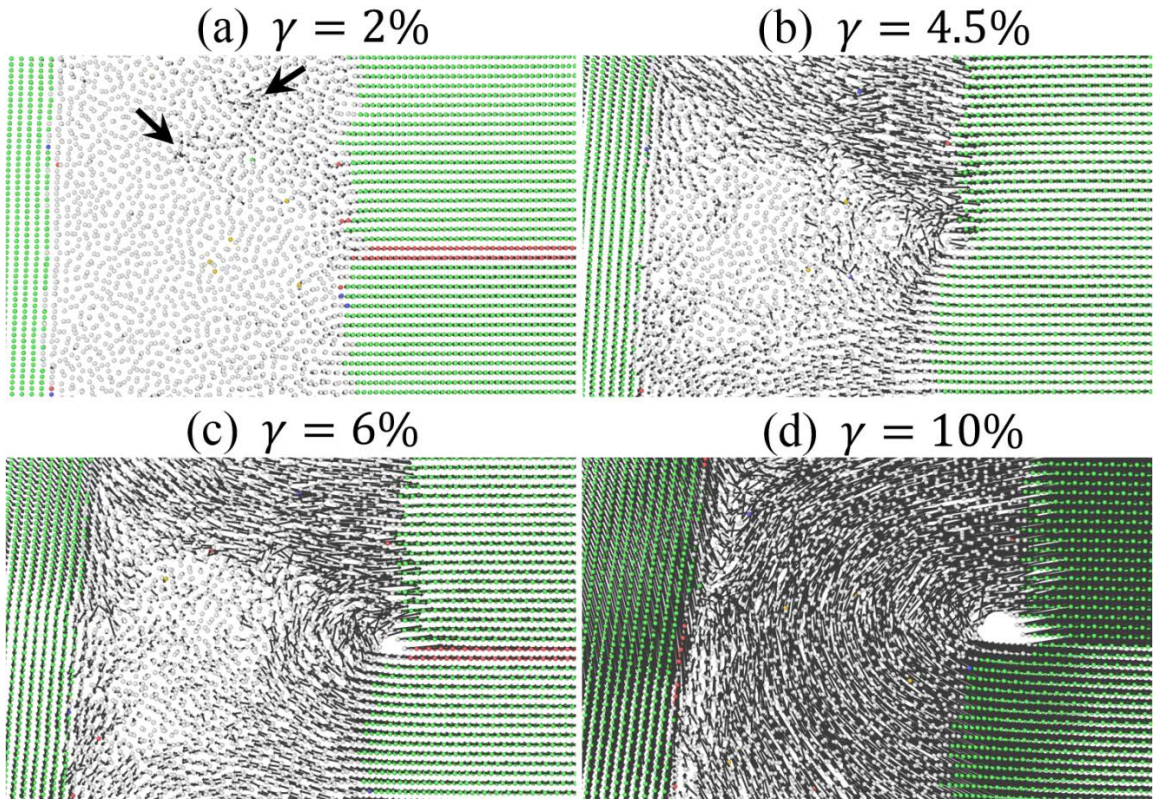

**Figure 9. The displacement field of a sample with 5.7 nm thick AIFs, loaded at shear strain rate of $10^8$ s$^{-1}$, at applied shear strains of (a) 2%, (b) 4.5%, (c) 6%, and (d) 10%. Atoms are colored according to CNA. The magnitude and direction of the displacement of each atom is indicated by the length and direction of the associated arrow. The two black arrows in (a) show the positions of two shear transformation zones.**

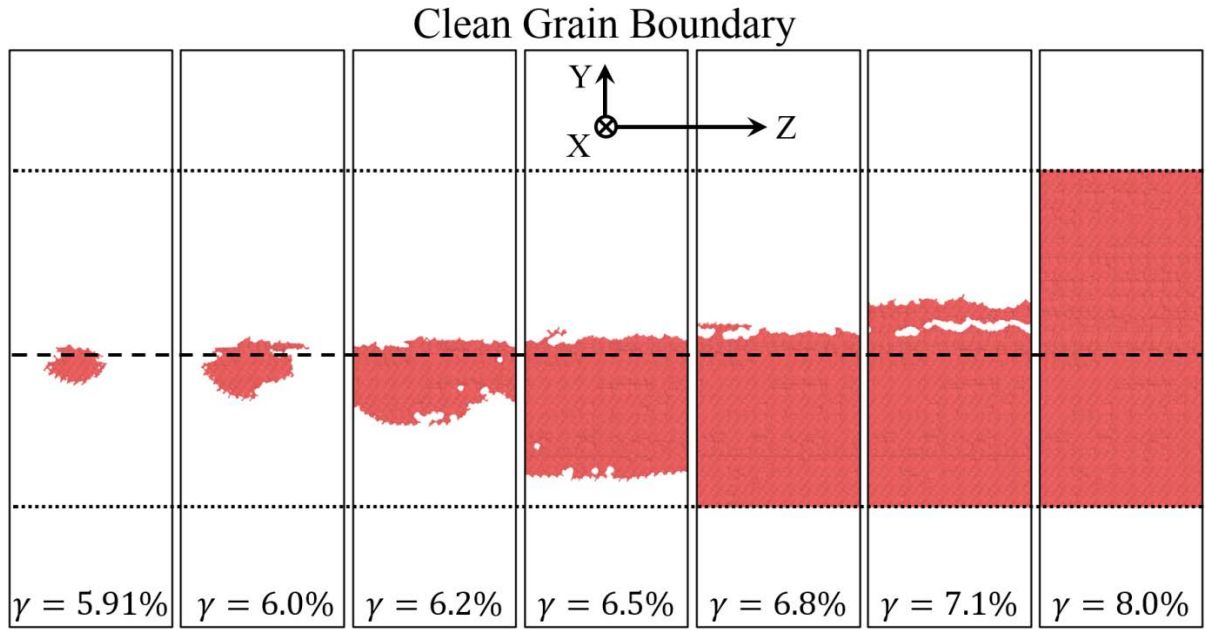

**Figure 10. Defect mesh points that sit at a crack surface show the shape evolution of the crack right after nucleation for a clean grain boundary. The cracks are viewed along the X-direction, with the dashed lines showing the position of the grain boundary-dislocation intersection and the dotted lines showing the boundaries of the meshes used to identify damage.**

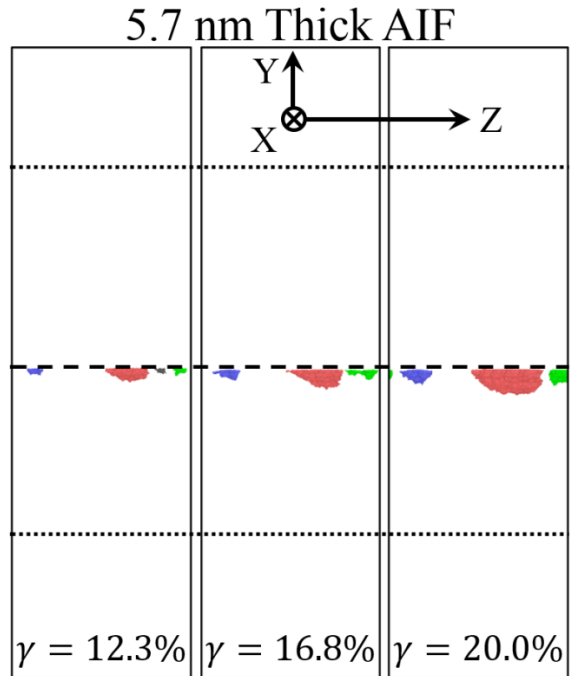

**Figure 11.** Defect mesh points that sit at crack surfaces show the shape evolution of the cracks right after nucleation for a 5.7 nm thick AIF. Different cracks are identified with different colors. The cracks are viewed along the X-direction, with the dashed lines showing the position of the ACI-dislocation intersection and the dotted lines showing the boundaries of the mesh used to identify damage.

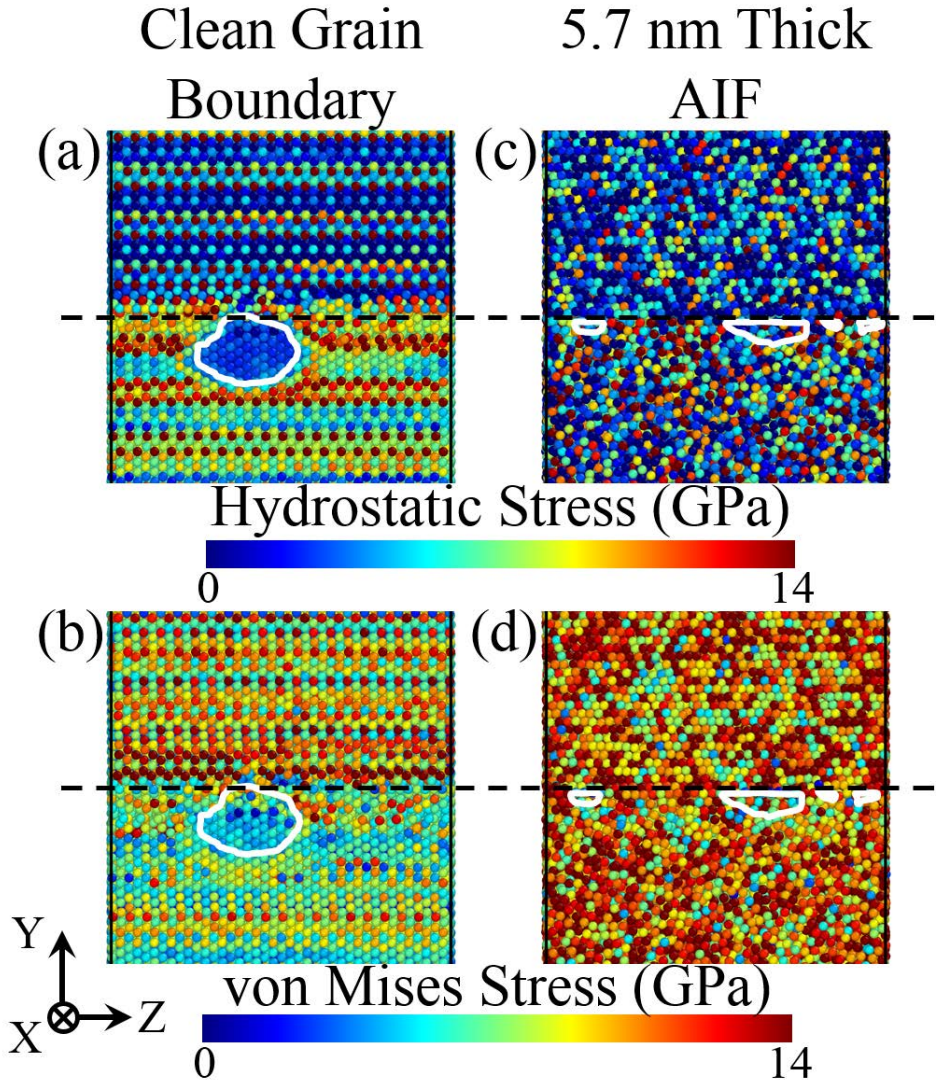

**Figure 12.** The distribution of (a) the hydrostatic stress and (b) the von Mises stress along the right grain boundary in Figure 10 and the distribution of (c) the hydrostatic stress and (d) the von Mises stress along the right ACI in Figure 11 right after crack nucleation. The viewing angle is the same as Figure 10 and 11, but only the vicinity of the dislocation-interface intersection is shown. The contours of the cracks are shown with solid white lines, while dashed black lines show the position of grain boundary/ACI-dislocation intersection.